\documentclass[aps,pre,reprint,showpacs,superscriptaddress]{revtex4-1}
\usepackage{amsmath}
\usepackage{amssymb}
\usepackage{gensymb}
\usepackage{graphicx}
\usepackage{epstopdf}
\begin{document}
	
	\title{Structural studies of the bond-orientational order and hexatic-smectic transition in liquid crystals of various compositions}
	
	\author{I.~A.~Zaluzhnyy}
	\affiliation{Deutsches Elektronen-Synchrotron DESY, Notkestra{\ss}e 85, D-22607 Hamburg, Germany}
	\affiliation{National Research Nuclear University MEPhI (Moscow Engineering Physics Institute), Kashirskoe shosse 31, 115409 Moscow, Russia}
	
	\author{R.~P.~Kurta}
	\affiliation{European XFEL GmbH, Holzkoppel 4, D-22869 Schenefeld, Germany}

	\author{E.~A.~Sulyanova}
	\affiliation{FSRC "Crystallography and Photonics", Russian Academy of Sciences, Leninskii prospect 59, 119333 Moscow, Russia}
	
	\author{O.~Yu.~Gorobtsov}
	\affiliation{Deutsches Elektronen-Synchrotron DESY, Notkestra{\ss}e 85, D-22607 Hamburg, Germany}
	
	\author{A.~G.~Shabalin}
	\affiliation{Deutsches Elektronen-Synchrotron DESY, Notkestra{\ss}e 85, D-22607 Hamburg, Germany}
	\affiliation{Present address: Department of Physics, University of California-San Diego, La Jolla, California 92093-0319, USA}
	
	\author{A.~V.~Zozulya}
	\affiliation{Deutsches Elektronen-Synchrotron DESY, Notkestra{\ss}e 85, D-22607 Hamburg, Germany}
	\affiliation{Present address: European XFEL GmbH, Holzkoppel 4, D-22869 Schenefeld, Germany}
	
	\author{A.~P.~Menushenkov}
	\affiliation{National Research Nuclear University MEPhI (Moscow Engineering Physics Institute), Kashirskoe shosse 31, 115409 Moscow, Russia}
	
	\author{M.~Sprung}
	\affiliation{Deutsches Elektronen-Synchrotron DESY, Notkestra{\ss}e 85, D-22607 Hamburg, Germany}
	
	\author{A.~Kr\'owczy\'{n}ski}
	\affiliation{Department of Chemistry, University of Warsaw, Zwirki I Wigury 101, 02-089, Warsaw, Poland}
	
	\author{E.~G\'orecka}
	\affiliation{Department of Chemistry, University of Warsaw, Zwirki I Wigury 101, 02-089, Warsaw, Poland}	
	
	\author{B.~I.~Ostrovskii}
	\email[Corresponding author: ]{ostenator@gmail.com}
	\affiliation{FSRC "Crystallography and Photonics", Russian Academy of Sciences, Leninskii prospect 59, 119333 Moscow, Russia}
	\affiliation{Landau Institute for Theoretical Physics, Russian Academy of Sciences, prospect akademika Semenova 1-A, 142432 Chernogolovka, Russia}
	
	\author{I.~A.~Vartanyants}
	\email[Corresponding author: ]{ivan.vartaniants@desy.de}
	\affiliation{Deutsches Elektronen-Synchrotron DESY, Notkestra{\ss}e 85, D-22607 Hamburg, Germany}
	\affiliation{National Research Nuclear University MEPhI (Moscow Engineering Physics Institute), Kashirskoe shosse 31, 115409 Moscow, Russia}

	\date{\today}
	
	\begin{abstract}
          We report on X-ray studies of freely suspended hexatic films of three different liquid crystal compounds. By applying angular X-ray cross-correlation analysis (XCCA) to the measured diffraction patterns the parameters of the bond-orientational (BO) order in the hexatic phase were directly determined. The temperature evolution of the BO order parameters was analyzed on the basis of the multicritical scaling theory (MCST). Our results confirmed the validity of the MCST in the whole temperature range of existence of the hexatic phase for all three compounds.  The temperature dependence of the BO order parameters in the vicinity of the hexatic-smectic transition was fitted by a conventional power law with a critical exponent $\beta\approx0.1$ of extremely small value. We found that the temperature dependence of higher order harmonics of the BO order scales as the powers of the first harmonic, with exponent equal to harmonic number. This indicates a nonlinear coupling of the BO order parameters of different order. It is shown that compounds of various composition, possessing different phase sequences, display the same thermodynamic behavior in the hexatic phase and in the vicinity of the smectic-hexatic phase transition.
	\end{abstract}
	
	\maketitle

\section{Introduction}
\label{sec1}
The hexatic phase was first predicted as an intermediate state between a crystal and a liquid in the theory of two-dimensional (2D) melting \cite{Halperin1978, Nelson2002}.
The 2D hexatic phase is characterized by a sixfold quasi-long-range bond-orientational (BO) order, while the positional order is short range. Phases with 2D hexatic order have been found in different systems, such as polymer colloids\cite{Murray1987,Kusner1994,Keim2007}, electrons at the surface of helium\cite{Glattli1988}, superconducting vortices \cite{Murray1990,Guillamon2009} and smectic liquid crystals (LCs) \cite{Pindak1981,Brock1986,Stoebe1995}.
In the latter case the hexatic phase was experimentally observed in a three-dimensional (3D) stack of the parallel molecular layers \cite{Pindak1981}.
The formation of the hexatic ordering in 3D can be hardly understood - according to Halperin and Nelson theory \cite{Halperin1978} the hexatic phase arises as a consequence of the broken translational symmetry of a 2D crystal, induced by dissociation of thermally excited dislocation pairs.
The transition to an isotropic liquid occurs only after a subsequent unbinding of disclination pairs.
Such a defect-mediated mechanism clearly does not work in 3D crystals due to difference in defect formation energy in 2D and 3D case.
Fortunately, the hexatic phase in LCs can be fully described on the basis of symmetry considerations, which does not imply any specific melting mechanism. For example, the quantitative characterization of the BO order in 2D and 3D hexatic LCs can be made using the multicritical scaling theory (MCST) developed by Aharony \textit{et al.} \cite{Aharony1986}.

The critical behavior of liquid crystals at the smectic-hexatic phase transition remains a challenging problem that is still not well understood. In the following we consider the smectic and hexatic phases in which elongated molecules are oriented perpendicular to the layers (Sm-A and Hex-B, respectively \cite{Jeu2003}).
The BO order in the Hex-B phase is characterized by a sixfold rotational symmetry and is described by the local ordering field
\begin{equation}
\label{Eq1}
\psi(\textbf{r})=|\psi(\textbf{r})|\exp[i6\theta(\textbf{r})] \, ,
\end{equation}
where  $\psi(\textbf{r})$ is the angle between the intermolecular bonds and some reference axis.
Since the BO order field is two-component, the Sm-A -- Hex-B phase transition belongs to the XY universality class, which is similar to superfluid helium or smectic-C liquid crystals.
Contrary to the expectations for XY critical behavior in 3D, the properties measured in the vicinity of the Sm-A -- Hex-B transition deviate markedly from these predictions.
For example, various experimental techniques yield an order parameter critical exponent $\beta=0.15-0.19$ \cite{Rosenblatt1982,Gorecka1994} and specific heat data for the most carefully studied Hex-B compound 65OBC yield the exponent $\alpha=0.6-0.65$ \cite{Huang1981,Haga1997,VanRoie2005,Mercuri2013}.
These values differ significantly from theoretical predictions for both the 3D XY critical exponents ($\alpha \approx -0.01$ and $\beta \approx 1/3$ \cite{Stanley1987,Pelissetto2002}) and a tricritical point ($\alpha=1/2$ and $\beta=1/4$).

The structure of hexatics is traditionally studied by means of X-ray or electron diffraction in a single-domain area of a hexatic film \cite{Pindak1981,Brock1986}.
Within such an approach the BO order parameters are determined by fitting the measured azimuthal intensity distribution with the Fourier cosine series.
In our previous X-ray studies \cite{Kurta2013,Zaluzhnyy2015} we have applied the angular X-ray cross-correlation analysis (XCCA) to study the BO order in single-domain Hex-B films.
Angular XCCA has been developed to investigate the structure of partially ordered systems to detect hidden symmetries and weak angular correlations \cite{Wochner2009,Altarelli2010,Kurta2013a,Kurta2015,Schroer2015,Kurta2016}.
These include various complex fluids, polymers, colloids, suspensions of biological molecules, block copolymers and liquid crystals.
In relation to the hexatic phase, XCCA enabled a direct determination of the sixfold BO order parameters from the ensemble of measured diffraction patterns \cite{Kurta2013,Zaluzhnyy2015}.
The application of XCCA allowed us to avoid the uncertainty of a fitting procedure and determine multiple harmonics of the BO order with great accuracy.
Furthermore, having the values of the BO order parameters one can reconstruct the angular resolved pair distribution function (PDF) and observe the formation of the hexatic phase in real space \cite{Zaluzhnyy2016}.

In the present work we investigated the structure of three different LC compounds, prepared as thick freely suspended hexatic films, which are truly 3D systems.
There are several motivations for this study.
First, we wanted to make sure that the character of the BO ordering in the Hex-B phase is the same for LC molecules of different compositions and various phase sequences.
Two of our LC materials, 75OBC and 3(10)OBC studied earlier \cite{Kurta2013, Zaluzhnyy2015, Zaluzhnyy2017} belong to the homologous series nmOBC and show a low-temperature 3D crystalline phase with a rectangular in-plane lattice possessing a herringbone order (so-called Cr-E phase) \cite{Jeu2003,Geer1993}.
The third compound, PIRO6, possesses very different chemical structure (with hydrogen bonds) and shows a low-temperature crystalline phase with a hexagonal in-plane lattice (Cr-B) \cite{Pyzuk1995}.
The latter is especially important on account of speculations on the possible influence of a low-temperature herringbone order on the properties of hexatics \cite{Bruinsma1982}.
Second, we studied the temperature dependence of the BO order parameters in the vicinity of the Sm-A -- Hex-B phase transition that allowed us to obtain and compare the values of the critical exponent $\beta$ for different compounds.
And finally, we performed analysis of the BO order parameters in the framework of MCST and proved the validity of this theory for the studied compounds.

This work is organized in the following way.
In the next section we discuss various theoretical approaches to describe the 2D and 3D hexatic ordering in LCs.
In the third section the basics of the angular X-ray cross-correlation is presented.
The fourth section contains the description of the experiment and properties of the studied compounds.
In the fifth section we present analysis of the positional order and the results of the spatially resolved studies of the single-domain formation in hexatic films.
Then we show the data on the BO order parameters and the results of application of the MCST to all three LC compounds.
The section completes with the analysis of the critical behavior of the BO order parameters in the vicinity of the Sm-A -- Hex-B transition.
The final section gives the concluding discussion.

\section{Theoretical aspects of the hexatic ordering}
\label{sec2}
\subsection{Stacked hexatic phase}
\label{sec2.1}
The in-plane structure in the Sm-A phase is liquid-like with
the positional correlations between molecules decaying exponentially.
Upon decreasing temperature a transition into the Hex-B phase, which displays the BO order, may occur.
This means that within smectic layers orientations of local hexagons persist over macroscopic distances, even in the absence of positional order.
For smectic layers this leads to a sixfold rotational symmetry and BO correlations are described in terms of the two-component ordering field introduced in the preceding section.
The above picture applies to the individual smectic layers that are 2D hexatics in nature.
It was experimentally confirmed that hexatic films consisted of one or two molecular layers represent true 2D systems \cite{Cheng1988,Chou1996,Chou1997}.
Birgeneau and Litster \cite{Birgeneau1978} suggested that some of the ordered smectic phases might actually be the 3D analogue of the hexatic phase described by Halperin and Nelson \cite{Halperin1978}.
In such case the coupling of layers in the third dimension introduces additional angular correlations from layer to layer.
Here the mean-square fluctuations of the bond angles remain finite in contrast to the situation in two dimensions, so a 3D (or stacked) hexatic phase exhibits true long-range BO order.
At the same time the in-plane positional correlations remain short-range and decay exponentially with a distance.

The in-plane structure factor in the Sm-A phase has the form of a broad ring due to the short-range positional correlations between molecules.
The presence of the BO order in the hexatic phase breaks the angular isotropy of the structure factor and leads to a sixfold modulation of the in-plane scattering.
A Fourier expansion of the azimuthal scattering profile gives \cite{Brock1986}
\begin{equation}
\label{Eq2}
I(\phi)=I_0\Bigg[\frac{1}{2}+\sum_{m=1}^{\infty}C_{6m}\cos\big(6m(\phi-\phi_0)\big)\Bigg] \, ,
\end{equation}
where $I_0$ is the angular-averaged scattered intensity, $\phi$ is the angle measured along the arc of the diffraction pattern, and $\phi_0$ corresponds to the orientation of the diffraction pattern.
The coefficients $C_{6m}$ can be considered as order parameters that measure the degree of the sixfold bond ordering in the hexatics \cite{Aharony1986,Aeppli1984}.
In addition, the hexatic phase exhibits a diffuse scattering profile in the radial direction, corresponding to the short-range positional order.
On cooling, the width of the radial intensity peak decreases simultaneously with a further development of the BO order.
This indicates a coupling between the positional correlations and the BO order \cite{Bruinsma1981}.
At certain stage the increase of the positional order leads to formation of the 2D in-plane lattices that are locked together into 3D crystalline structures.
These phases possess either a hexagonal in-plane crystal lattice (Cr-B), or a rectangular lattice with the herringbone order (Cr-E).

\subsection{In-plane positional order in hexatics}
\label{sec2.2}
Due to interplay between the liquid density fluctuations and the BO order parameters analysis of the radial intensity profiles in the hexatic phase is a complicated task \cite{Aeppli1984,Bruinsma1981}.
The X-ray scattering technique measures the Fourier-transformed density-density autocorrelation function.
In fact, the measurement of the evolution of the BO order is possible only due to coupling between the liquid density fluctuations and the hexatic order parameters.
The change of the liquid structure factor (LSF) in the vicinity of the hexatic-smectic transition was studied in the framework of the phenomenological XY model by Aeppli and Bruinsma \cite{Aeppli1984}.
The authors have shown that the form of LSF (measured as a radial scan through one of the sixfold diffraction maxima) depends strongly on the mean square amplitude of fluctuations of the BO order parameters.
In the Sm-A phase at high temperatures and in the Hex-B phase at low temperatures far away from the phase transition temperature $T_C$ the LSF can be described by the Lorentzian profile with a half-width $\gamma$
\begin{equation}
\label{Eq3}
S(q)\propto\frac{\gamma^2}{(q-q_0)^2+\gamma^2}\, ,
\end{equation}
centered at the preferred in-plane wave vector $q_0\approx4\pi/a\sqrt{3}$, where $a$ is the average lateral molecular separation. In this case the positional correlations decay as $\sim r^{-1/2} \exp⁡(-\gamma r)$ for 2D systems at large distances \cite{Jeu2003}.

The most interesting behavior occurs in the vicinity of the smectic-hexatic transition, where the fluctuations of the BO field dominate.
According to the theoretical predictions \cite{Aeppli1984} the LSF is well approximated in this region by the square root of a Lorentzian (SRL)
\begin{equation}
\label{Eq4}
S(q)\propto\frac{\gamma}{\sqrt{(q-q_0)^2+\gamma^2}}\, ,
\end{equation}
where the value of $\gamma$ is determined by the mean square amplitude of the angular fluctuations $\langle|\psi|^2\rangle$.
For such a profile the positional correlations in 2D systems decay as $\sim r^{-1} \exp⁡(-\gamma r)$ \cite{Jeu2003}.
In this work we defined the positional correlation length, that characterizes the exponential decay of positional correlations between the molecules as $\xi=1/\Delta q$, where $\Delta q$ is a half width at half maximum (HWHM) of the diffraction peak in the radial direction.
If the diffraction peak is approximated by the Lorentzian function, $\xi=1/\gamma$, and in the case of the SRL function, $\xi=1/\sqrt{3}\gamma$.

As a consequence of a functional dependence of the diffraction peak shape on the mean square amplitude $\langle|\psi|^2\rangle$, the correlation length at the smectic-hexatic transition starts to grow in the high temperature smectic phase, shows the maximum of the derivative $d\xi/d$T at the phase transition point $T_C$, and then nearly saturates deep in the hexatic phase.
This corresponds to the presence of an inflection point on the temperature dependence of the correlation length $\xi$ at the phase transition temperature $T_C$.
Interestingly, similar temperature dependence with the inflection point at $T_C$ is predicted also for the peaks maximum position $q_0$ \cite{Aeppli1984}.
Such behavior has been observed experimentally for many hexatic LCs \cite{Jeu2003,Kurta2013,Davey1984}.
The temperature width of the region where the BO order fluctuations are really strong has not been evaluated in the theory \cite{Aeppli1984}.
From the X-ray experiments it can be estimated as few degrees, in accordance with the range of the divergent behavior of the specific-heat at the Hex-B -- Sm-A phase transition \cite{Huang1981,Haga1997,VanRoie2005,Mercuri2013}.

\subsection{Multicritical scaling theory}
\label{sec2.3}
\begin{figure}
	\centering
	\includegraphics[height=6cm]{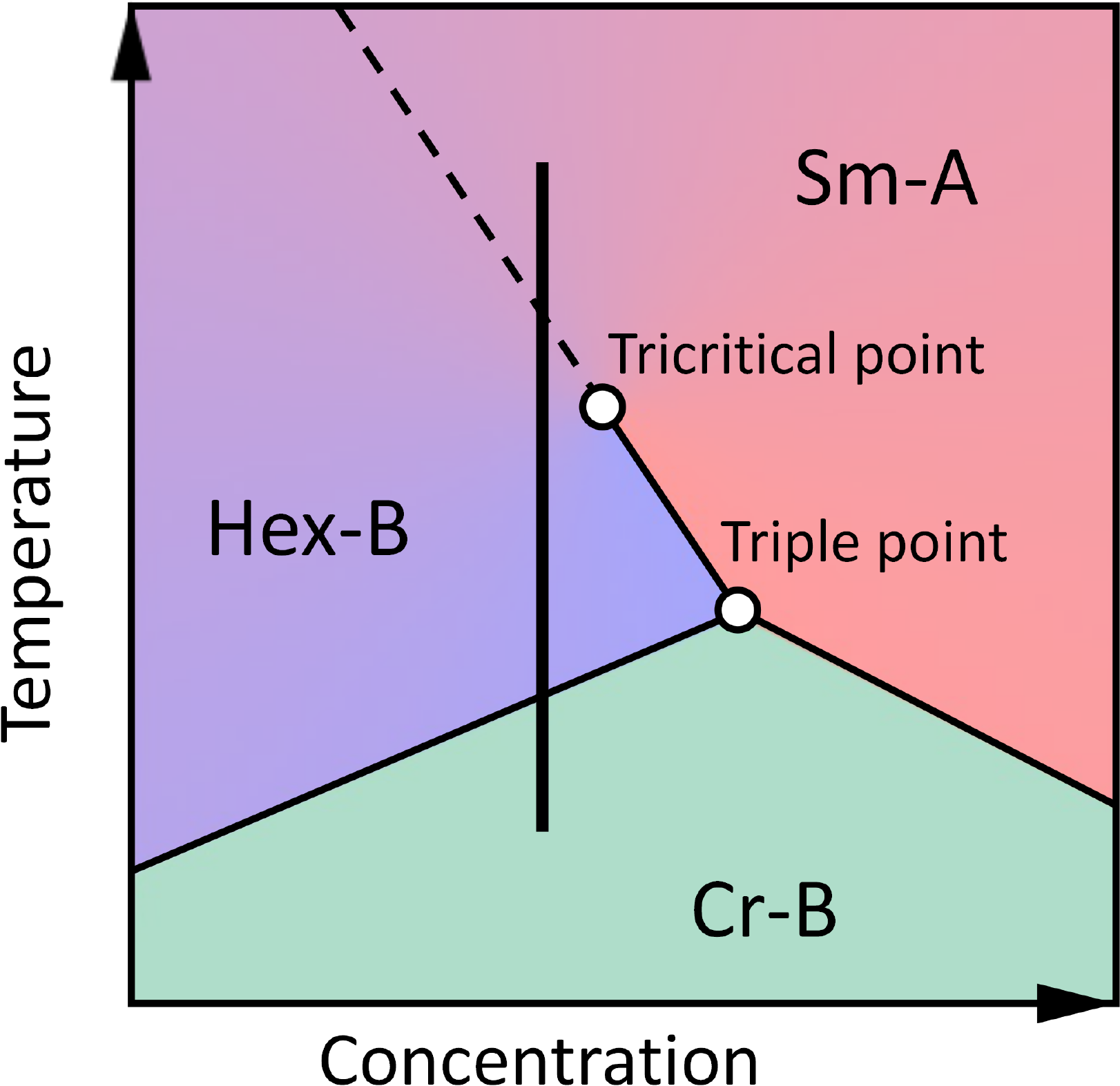}
	\caption{Generic temperature-concentration phase diagram, representing phase transitions between Sm-A, Hex-B and Cr-B phases of LCs. Solid lines denotes first-order transition and dash line corresponds to second-order transition. Black vertical line represents the phase sequence for certain compounds.}
	\label{Fig1}
\end{figure}

In the absence of a microscopic theory of the hexatic ordering in the stacked (3D) smectics we consistently use phenomenological theories, which are based on symmetry considerations and do not rely on any melting mechanism \cite{Aharony1986,Aeppli1984}.
The ordering field of the Sm-A -- Hex-B transition is a two-component BO order parameter.
This places this transition in the XY universality class and allows it to be continuous in 3D.
General thermodynamic arguments allow to reconstruct the topology of the hexatic phase diagram in the vicinity of the Sm-A -- Sm-B -- Cr-B triple point. A generic phase diagram suggested by Aharony \textit{et al.} \cite{Aharony1986} is shown in Fig.~\ref{Fig1}.
The Sm-A -- Cr-B transition in three dimensions is always first-order due to the long-range positional order and the presence of a cubic term in the free-energy expansion in powers of the Fourier coefficients of the density \cite{Stoebe1995a}.
The same cubic term is present for a Hex-B -- Cr-B transition and gives rise to a first-order transition in 3D.
These two first-order lines meet at a triple point with different slopes, since the coupling between the BO order parameter and the amplitudes of the density waves shifts the transition Hex-B -- Cr-B relative to the continuation of the Sm-A -- Cr-B line.
Such discontinuity in the slope at the triple point causes the Sm-A -- Hex-B transition to become first order in its vicinity \cite{Aharony1986}.
This introduces a tricritical point on the Sm-A -- Hex-B phase transition line.
There are also other interactions which potentially can shift the Sm-A -- Hex-B phase transition to the tricritical region.
These include the coupling of the BO order with the short-range herringbone molecular packings \cite{Bruinsma1982}, interaction of BO order with in-plane positional strains \cite{Haga1997}, and coupling between the hexatic order and layer displacement fluctuations \cite{Selinger1988}.
The precise adiabatic scanning calorimetry study of the hexatic compound 65OBC by Van Roie \textit{et al}. \cite{VanRoie2005} have proved that the Hex-B - Sm-A transition is indeed very weakly first order \cite{Haga1997,Mercuri2013}.
The latent heat at the transition was found to be as small as $\Delta H\approx0.04\ \text{J/g}$.

Assuming that Sm-A -- Hex-B phase transition lies in a crossover region between mean field (tricritial behavior) and 3D XY universality class, Aharony \textit{et al.} \cite{Brock1986,Aharony1986} have shown that the BO parameters $C_{6m}$ defined in Eq.~(\ref{Eq2}) in the frame of MCST satisfy the following scaling law
\begin{equation}
\label{Eq5}
C_{6m}=(C_{6})^{\sigma_m} \, ,
\end{equation}
with the exponent $\sigma_m$ of the form
\begin{equation}
\label{Eq6}
\sigma_m=m+x_m\cdot m\cdot(m-1) \, .
\end{equation}
Here the parameter $x_m$ also depends on the order $m$ and can be expanded into a series in powers of $m$:
\begin{equation}
\label{Eq7}
x_m\simeq\lambda-\mu m + \nu m^2 - \ldots\, .
\end{equation}
Two- and three-dimensional hexatic behavior in smectic films can be distinguished by the values of the expansion coefficient $\lambda$.
Theory predicts \cite{Aharony1986,Aharony1988,Paczuski1988}. that it is equal to $\lambda=0.3$ and $\lambda=1$ in the 3D and 2D cases, respectively.
These predictions were in a good agreement with the experimental works \cite{Brock1986,Cheng1988,Chou1996}.
Theoretically determined next correction term $\mu$ is two orders of magnitude smaller and equals approximately to 0.008 in the 3D case \cite{Aharony1986}.

\section{X-ray cross-correlation analysis}
\label{sec3}
XCCA allows one to study the BO order by analysing of the angular correlations of the scattered intensity \cite{Altarelli2010,Kurta2013,Kurta2016,Kurta2012}.
A two-point angular cross-correlation function (CCF) of intensity is defined as
\begin{equation}
\label{Eq8}
G(q,\Delta)=\langle I(q,\phi)I(q,\phi+\Delta)\rangle_\phi \, .
\end{equation}
Here $(q,\phi)$ are the polar coordinates in the 2D detector plane, $\Delta$ is the angular coordinate, and $\langle\ldots\rangle_\phi$ denotes the angular average around a ring of scattering of a radius $q$ (see Fig.~\ref{Fig2}). Since the CCF is an even function, it can be expanded into a cosine Fourier series
\begin{equation}
\label{Eq9}
G(q,\Delta)=G_0(q)+2\sum_{n=1}^{\infty}G_n(q)\cos(n\Delta)\, ,
\end{equation}
where the Fourier coefficients (FCs) are
\begin{equation}
\label{Eq10}
G_n(q)=\frac{1}{2\pi}\int_{0}^{2\pi}G(q,\Delta)\cos(n\Delta)\mathrm{d}\Delta\,.
\end{equation}
These Fourier components of the CCF are directly related to the corresponding complex valued Fourier components of the scattered intensity
\begin{equation}
\label{Eq11}
G_n(q)=|I_n(q)|^2 \, ,
\end{equation}
where
\begin{equation}
\label{Eq12}
I_n(q)=|I_n(q)|\exp(i\phi_n(q))=\frac{1}{2\pi}\int_{0}^{2\pi}I(q,\phi)e^{-in\phi}\mathrm{d}\phi \, .
\end{equation}
Here $|I_n(q)|$ and $\phi_n(q)$ are the magnitude and phase of the $n$-th FC of intensity and $I_0 (q)$ is the intensity averaged over a scattering ring of a radius $q$.

A large advantage of using the FCs $G_n (q)$ is that they can be averaged over all measured diffraction patterns
\begin{equation}
\label{Eq13}
\langle G_n(q) \rangle_P=\frac{1}{P}\sum_{p=1}^{P}G_n^p(q),
\end{equation}
where the index $p$ enumerates diffraction patterns and $P$ is a total number of measured diffraction patterns.
When a focused X-ray beam is used the local probe is much smaller than the length scale of orientational order.
In this case the averaged FCs  $\langle G_n (q)\rangle_P$ are related to the Fourier components of intensity \cite{Kurta2013a}
\begin{equation}
\label{Eq14}
\langle G_n(q) \rangle_P=|I_n(q)|^2\, ,
\end{equation}
where intensity $I_n (q)$ refers to the scattering from such a region with uniform orientation.

\section{Experiment}
\label{sec4}
\subsection{Samples}
\label{sec4.1}
We have used three different LC compounds that all exhibit the Hex-B phase within a certain temperature range.
At higher temperatures all compounds show the Sm-A phase.
The chemical structure and the transition temperatures for LC compounds under investigation are shown in Table~\ref{Table1}.
Two LC materials 75OBC and 3(10)OBC belong to the homologous series nmOBC \cite{Jeu2003,Geer1993}.
The 75OBC exhibits the Hex-B phase in the range \cite{Kurta2013,Geer1993,Geer1991} of 59 -- 63.8~$^\circ$C, while the 3(10)OBC shows the Hex-B phase in much wider range of temperatures 54 -- 66.3~$^\circ$C \cite{Zaluzhnyy2015,Stoebe1992}.
Both mesogens exhibit a low-temperature crystalline phase with a rectangular in-plane lattice possessing a herringbone order (Cr-E phase \cite{Jeu2003,Geer1993,Geer1991}).
The third compound, PIRO6, possesses a very different chemical structure: its aromatic part consists of three rings, the central of which contains a weak hydrogen bond (see Table~\ref{Table1}).
The Hex-B phase exists in PIRO6 in the temperature range from 85 to 92.6 $^\circ$C; contrary to the first two mesogens PIRO6  shows a low-temperature crystalline phase with a hexagonal in-plane lattice (Cr-B) \cite{Pyzuk1995}. Below we use relative temperature $\Delta t=T-T_C$, where $T_C$ is the temperature of the Sm-A -- Hex-B phase transition for a given LC compound.

\begin{table*}
	\small
	\caption{Properties of the LC compounds}
	\label{Table1}
	\begin{tabular*}{\textwidth}{@{\extracolsep{\fill}}lp{75mm}lp{30mm}}
		\hline
		Compound & Chemical structure & Sm-A -- Hex-B & Hex-B -- Cr \\
		\hline
    	3(10)OBC & \includegraphics{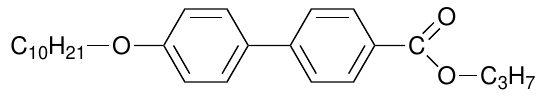}  \newline \textit{n}-propyl-4$'$-\textit{n}-decyloxybephenyl-4-carboxylate
		& $T_C\approx66.3\ ^\circ\text{C}$ & $T_{Cr}\approx54.0\ ^\circ\text{C}$ \newline Cr-E (orthorhombic)\\
		75OBC &\includegraphics{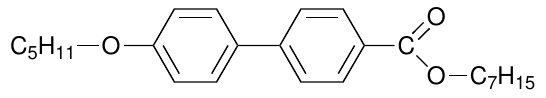} \newline \textit{n}-heptyl-4$'$-\textit{n}-pentyloxybephenyl-4-carboxylate
		& $T_C\approx63.8\ ^\circ\text{C}$ & $T_{Cr}\approx59.0\ ^\circ\text{C}$ \newline Cr-E (orthorhombic) \\
		PIRO6 &\includegraphics{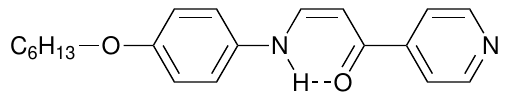} \newline 1-(4$'$ pirydyl)-3-(4-hexyloxyphyenyloamine)-prop-2-en-1-on
		& $T_C\approx92.6\ ^\circ\text{C}$ & $T_{Cr}\approx86.0\ ^\circ\text{C}$ \newline Cr-B (hexagonal) \\
		\hline
	\end{tabular*}
\end{table*}

\subsection{Experimental setup}
\label{sec4.2}
\begin{figure}
	\centering
	\includegraphics[width=8.3cm]{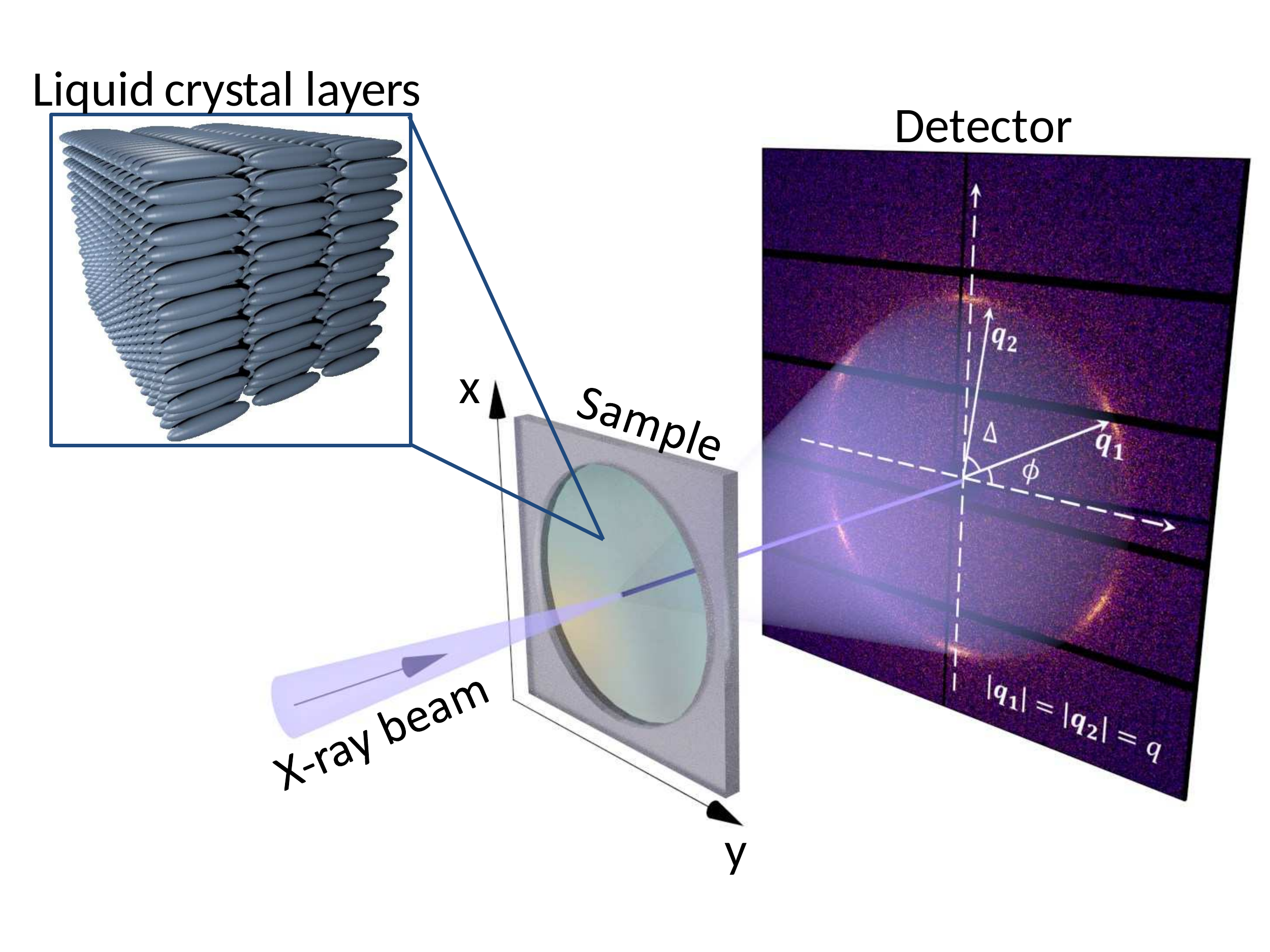}
	\caption{Scheme of the experimental setup. An X-ray beam focused by CRLs is incoming perpendicular to the surface of a freely suspended LC film. The diffraction pattern in transmission geometry is measured by a 2D detector positioned behind a sample. In the inset the arrangement of elongated LC molecules in the smectic (hexatic) planes is shown. In this geometry the measured diffraction pattern corresponds to the in-layer structure of LC.}
	\label{Fig2}
\end{figure}

The experiments were conducted at the coherence beamline P10 of the PETRA III synchrotron source at DESY.
The experimental setup is schematically shown in Fig.~\ref{Fig2}.
The energy of incident X-ray photons was 13 keV that corresponds to the wavelength of $\lambda=0.0954$ nm.
The X-ray beam was focused by the set of compound refractive lenses (CRLs) to the size of $2\times3$ $\mu$m$^2$ (vertical vs. horizontal) at full width at half maximum (FWHM) with the flux about $3\times10^{10}$ photons/s.
To collect the scattering signal at the diffraction angle $2\theta\approx12^{\circ}$ ($q \approx 14$ nm$^{-1}$, $q = 4\pi/\lambda \sin \theta$) the 2D detector Pilatus 1M ($981\times1043$ pixels of $172\times172$ $\mu$m$^2$ size) was positioned perpendicular to the incident beam at the distance of 263 mm downstream from the sample.
At each temperature the sample was scanned in the plane perpendicular to the incident beam direction with 11 $\mu$m step size to analyse the spatial variations of the BO order.

In our experiments we have used thick freely suspended hexatic films that are especially suitable for an X-ray scattering study \cite{Jeu2003}.
These films have a controlled thickness; the films are not influenced by substrate interactions and their two surfaces induce an almost perfect 2D alignment of the smectic layers.
The thickness of the films was measured by AVANTES fiber optical spectrometer and was in the range of 3-5 $\mu$m for 75OBC and 3(10)OBC samples and about 15 $\mu$m for PIRO6 sample (thousands of molecular layers).
Freely suspended LC films were drawn across a circular glass aperture of 2 mm in diameter inside the FS1 temperature stage from INSTEC at about 10 $^\circ$C above the temperature of the Sm-A -- Hex-B phase transition.
The sample stage was aligned in such a way that the surfaces of the LC films were perpendicular to the incident beam.
During the experiment the sample was gradually cooled with a temperature ramp of 0.05-0.1 $^\circ$C/min to observe structural changes at the Sm-A -- Hex-B phase transition point and in the hexatic phase.
Temperature was monitored by a mK1000 temperature controller with an accuracy of 0.005 $^\circ$C.
At each temperature a large number of diffraction patterns were collected for better statistics.
The exposure time was chosen in the range of 0.5 -- 0.6 s to avoid radiation damage of the samples.

\begin{figure}
	\centering
	\includegraphics[width=8.3cm]{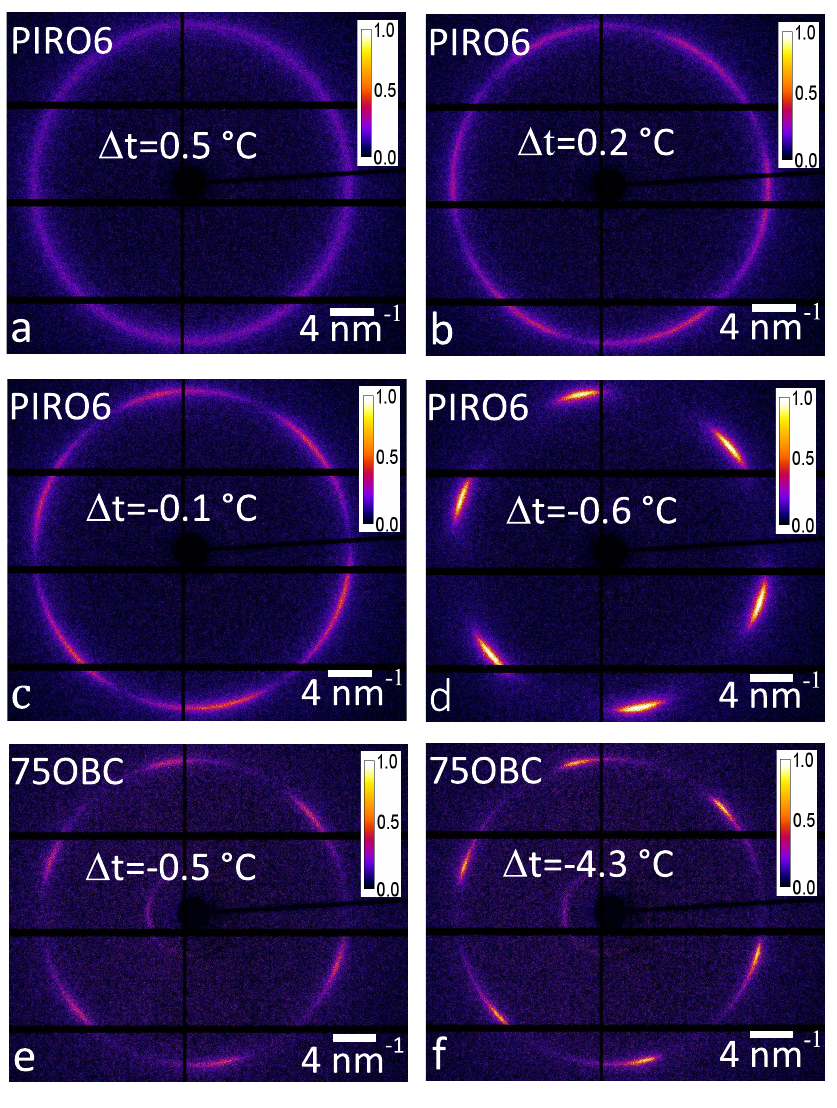}
	\caption{Diffraction patterns collected at different temperatures in PIRO6 and 75OBC samples. (a)  Sm-A phase, (b) Sm-A -- Hex-B phase transition region, (c,d) Hex-B phase of the PIRO6 sample, and (e,f) Hex-B phase of the 75OBC sample. Black stripes correspond to the detector gaps where no scattering signal was measured.}
	\label{Fig3}
\end{figure}

Typical diffraction patterns for PIRO6 and 75OBC samples at different temperatures are shown in Fig.~\ref{Fig3}(a-f).
In the Sm-A phase one can see a uniform ring that corresponds to the liquid-like structure of smectic layers (Fig.~\ref{Fig3}(a)). On cooling the intensity modulation appears along the scattering ring thus indicating the formation of the hexatic order within the layers (Fig.~\ref{Fig3}(b-c)). At lower temperatures the scattering ring splits into six distinct arcs that correspond to the increase of the BO order (Fig.~\ref{Fig3}(d)).

\section{Results and disscussion}
\label{sec5}
\subsection{The in-plane positional order}
\label{sec5.1}
\begin{figure*}
	\centering
	\includegraphics[width=17.1 cm]{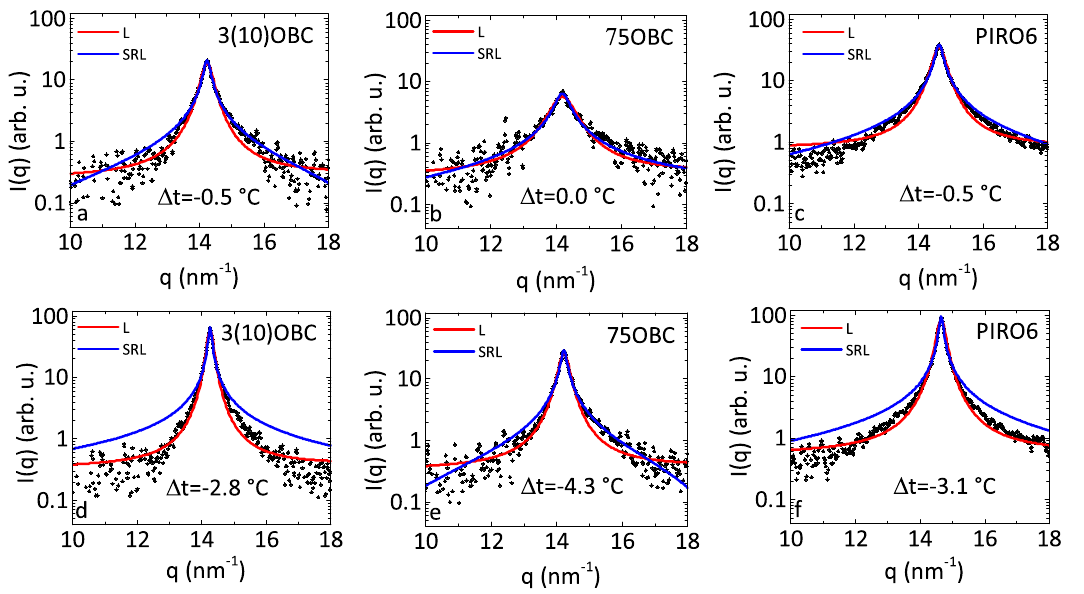}
	\caption{Intensity distribution $I(q)$ along the radial direction through the diffraction peak in the vicinity of the Sm-A -- Hex-B phase transition (a-c) and at lower temperatures in the Hex-B phase (d-f) for different compounds. The profiles are shown in semi-logarithmic scale. Black circles represent the experimental data, red and blue lines show fits with the Lorentzian (L) and SRL functions, respectively.}
	\label{Fig4}
\end{figure*}

In the present work we studied the in-plane positional order in the Sm-A and Hex-B phases using the shape analysis of the radial cross-sections of the diffraction patterns.
Since the molecular form factor slowly varies with $q$, we assumed that the shape of the diffraction peaks is governed by the structure factor.
Thus, the intensity distribution $I(q)$ in the radial direction through the maximum of one of the sixfold diffraction peaks in the Hex-B phase was averaged over different positions in the sample and then fitted with the Lorentzian  and SRL functions (see Section~\ref{sec2.2} for theoretical discussion).
The corresponding intensity profiles for all three compounds in the vicinity of the Sm-A -- Hex-B phase transition and at lower temperatures in the Hex-B phase after background subtraction are shown in Fig.~\ref{Fig4}.
For the case of 3(10)OBC and PIRO6 samples the SRL function gives better approximation for $I(q)$ in the temperature interval $-0.8\ ^\circ\text{C}\le\Delta t\le0\ ^\circ\text{C}$ and $-2.7\ ^\circ\text{C}\le\Delta t\le0.3\ ^\circ\text{C}$, respectively, that is in the close vicinity of the Sm-A -- Hex-B phase transition temperature (Fig.~\ref{Fig4}(a,c)). Out of this range the Lorentzian function fits the experimental data better (Fig.~\ref{Fig4}(d,f)) \cite{Zaluzhnyy2015}.
For the 75OBC sample the quality of the data does not allow us unambiguously distinguish between L and SRL fits  (Fig.~\ref{Fig4}(b,e)).

\begin{figure}
	\centering
	\includegraphics[width=8.3 cm]{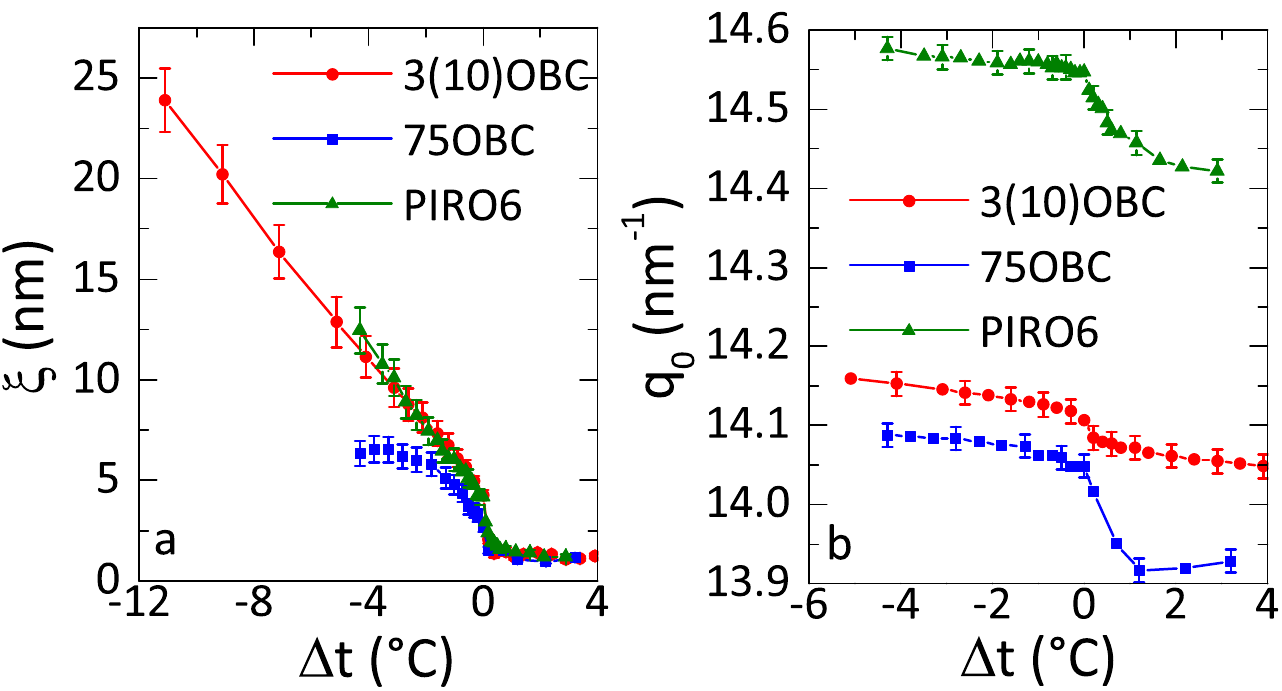}
	\caption{(a) Temperature dependence of the positional correlation length~$\xi$; (b) Temperature dependence of the diffraction peak maximum position $q_0$ for three different compounds.}
	\label{Fig5}
\end{figure}

In Fig.~\ref{Fig5}(a) the positional correlation length $\xi=1/\Delta q$ is shown for all three samples as a function of a relative temperature $\Delta t$.
In the Sm-A phase $\xi$ is about 1-2 nm and does not change with the temperature. On cooling, upon approaching the Hex-B phase the correlation length $\xi$ starts to increase and shows the maximum of the derivative $d\xi/dT$ at the phase transition point $T_C$.
This indicates a coupling between the spatial density fluctuations and the BO order.
The temperature dependence of the positional correlation length differs from one hexatic compound to another due to difference in the value of a coupling constant of above interaction.
On cooling, the correlation length $\xi$ either saturates deeper in the Hex-B phase (75OBC) or linearly increases as happens in 3(10)OBC and PIRO6 hexatic films.
Close to the crystallization temperature the in-plane positional correlation length increases by about an order of magnitude (see Table~\ref{Table2}). The same trend was observed for other LC compounds, such as 46OBC \cite{Davey1984} or surface hexatic layers in 4O.8 films \cite{Jeu2003a}, where $\xi$ reaches the value of about 16 nm and 40 nm, respectively.

The peak position $q_0$ determined by the average in-plane distance between LC molecules, slightly increases upon the temperature decrease (Fig.~\ref{Fig5}(b)).
For all three compounds  we detected the presence of the inflection point in the peak position $q_0$  temperature dependence at the Sm-A -- Hex-B phase transition temperature.
This finding is in accordance with the theory of Aeppli and Bruinsma \cite{Aeppli1984}, which predicts such behavior as a consequence of growing fluctuations of the BO order parameter \cite{Davey1984}.
These observations indicate that the position of the inflection point on the $q_0$ temperature dependence provides the most reliable determination of the temperature of the Sm-A -- Hex-B phase transition.

\begin{table}
	\small
	\caption{\ Values of different parameters obtained from the experiment}
	\label{Table2}
	\begin{tabular*}{0.5\textwidth}{@{\extracolsep{\fill}}llll}
		\hline
		Parameters & 3(10)OBC & 75OBC & PIRO6 \\
		\hline
		Max. positional \\ correlation length,\\ $\xi_{max}$ (nm) & $24\pm1.5$ & $6\pm1$ & $13\pm1$ \\
		Temperature range \\ of the hexatic phase \\ existence, $\Delta t$ ($^\circ$C) & 11 & 5 & 6.6 \\
		Max. number of the \\ BO order  parameters, \\ $m_{\text{max}}$ & 25 & 7 & 6 \\
		MCST parameters, & & & \\
		$\lambda$ & $0.31\pm0.015$ & $0.27\pm0.015$ & $0.29\pm0.015$ \\
		$\mu$ & $0.009\pm0.001$ & - & - \\
		$\nu$ & $\sim10^{-4}$ & - & - \\
		\hline
	\end{tabular*}
\end{table}

\subsection{Spatially resolved studies}
\label{sec5.2}

By scanning the sample with the incident X-ray beam, we were able to probe the spatial variation of the BO order within the plane of the Hex-B films \cite{Zaluzhnyy2015}.
To study the BO order we used angular Fourier decomposition of the scattered intensity at each position of the beam:
\begin{equation}
\label{Eq15}
I(q,\phi)=I_0(q)+2\sum_{n=1}^{\infty}|I_n(q)|\cos\left(n\phi+\phi_n(q)\right) \, ,
\end{equation}
where the magnitudes $|I_n (q)|$ and phases $\phi_n (q)$ are defined in Eq.~(\ref{Eq12}).
The FCs of the intensity were calculated along the ring of radius $q_0$.
We choose the magnitude of the dominant first harmonic $|I_6 (q_0)|$ of the sixfold FCs to characterize the degree of the BO order in the hexatic phase. At the same time the phase of the first harmonic $\phi_6 (q_0)$ determines the orientation of the molecular bonds.

\begin{figure}
	\centering
	\includegraphics[width=8.3 cm]{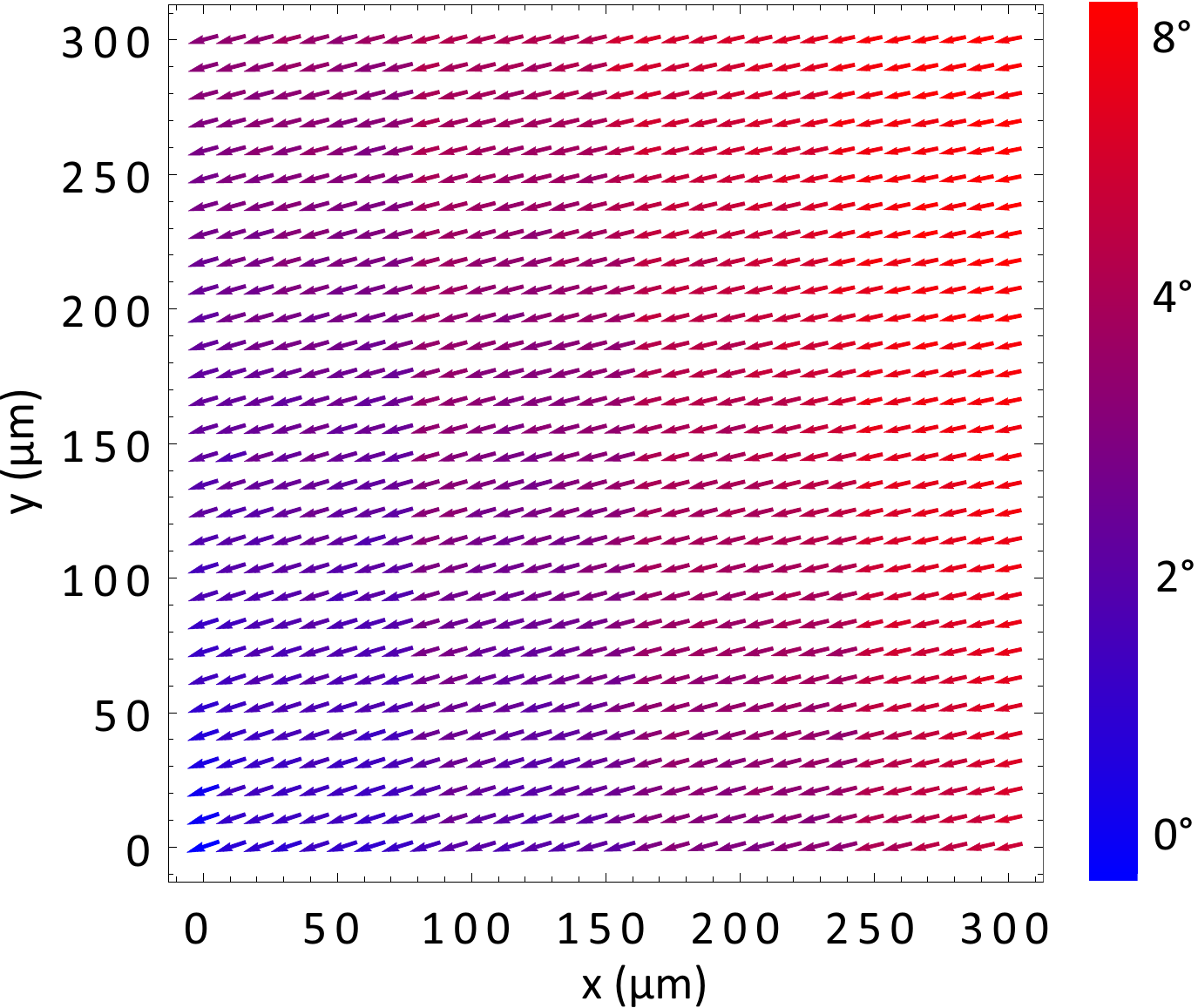}
	\caption{Spatially resolved 2D map of the 6th FC for the 3(10)OBC sample at the relative temperature $\Delta t=-5.3\ ^\circ°\text{C}$.  The length of a vector is proportional to the magnitude $|I_6 (q_0)|$ of the BO order parameter; the direction of a vector is associated with the phase $\phi_6 (q_0)$ of the first harmonic of the BO order. Color represents a relative change of the phase (in degrees).}
	\label{Fig6}
\end{figure}

In Fig.~\ref{Fig6} the spatial variation of $I_6 (q_0)$ is shown for the 3(10)OBC sample over large scanning area ($300\times300$ $\mu$m$^2$) at  a relative temperature $\Delta t=-5.3\ ^\circ\text{C}$.
In this 2D map the phase of the BO order parameter, i.e. the bond angle $\theta(\textbf{r})$, changes by approximately 7$^\circ$ at a distance of the order of 300 $\mu$m, while the magnitude $|I_6 (q_0)|$ is constant.
We expect that this is due to the occurrence of large scale hexatic single-domains with different bond angles over the area of the film.
The continuous variation of the phases of the BO order parameter within such domains allows the bond angles from different domains to match themselves tangentially.
This permits to avoid the formation of the borders between hexatic domains, which costs additional energy (of the order of $K_A(\partial\theta/\partial r)^2$, where $K_A$ is the orientational stiffness constant of the hexatic ordering field \cite{Aharony1988}).

The formation of large-scale single-domains of this type, as it is shown in Fig.~\ref{Fig6}, was typical for 3(10)OBC and PIRO6 samples where the diffraction peaks have symmetrical shape in azimuthal direction (see Fig.~\ref{Fig3}(a-d)).
This means that in all layers the bond angle $\theta(\textbf{r})$ has the same mean value, and the predominant orientation of intermolecular bonds is preserved in all layers.
A single domain with the uniform orientation is formed across the whole film.
It was not the case for the 75OBC sample, where the diffraction peaks mostly have asymmetrical shape in the azimuthal direction: for example, longer tails in the clockwise direction and steeper tails in the opposite direction (Fig.~\ref{Fig3}(e-f)).
The same asymmetry was observed at different spatial positions in the sample. We attribute an observed asymmetry of the diffraction peaks to a superposition of two or more domains in the illuminated area of the film with slightly different orientation of the intermolecular bonds.
The question arises whether these domains are originated from the surface and then penetrate due to angular BO interactions through the whole film, or whether these domains can be formed independently in the interior layers of the thick films.

It is known that in liquid crystals a free surface usually stabilizes a higher-ordered phase which is only observed at lower temperatures in the bulk \cite{Jeu2003}.
For example, transmission electron diffraction in thin 4O.8 films \cite{Chao1996} provided evidence of the following crystallization scenario: the hexagonal crystal grows into the smectic phase in a layer-by-layer fashion involving an intermediate hexatic phase.
At the same time, the direction of the hexatic (crystal) axes in the top layers at both sides of the film appear to orient in the same direction, though there are liquid smectic layers in between them.
Thus, it is clear that the hexatic domain starts to grow from the surface of the Sm-A film upon approaching the Sm-A -- Hex-B transition in bulk.
However, the surface epitaxy observed in the thin films might be lost in our thick (thousands layers) films.
As a result the bond-orientation of the hexatic domains growing from the top layers at both sides of the film can be different.
This necessary leads to smooth variation of the bond angle $\theta(\textbf{r})$ in the transient area between the surface.
From the above reasoning the formation of the single hexatic domains in the interior layers of the smectic film is unlikely.

\subsection{The bond-orientational order}
\label{sec5.3}
\begin{figure*}
	\centering
	\includegraphics[width=17.1 cm]{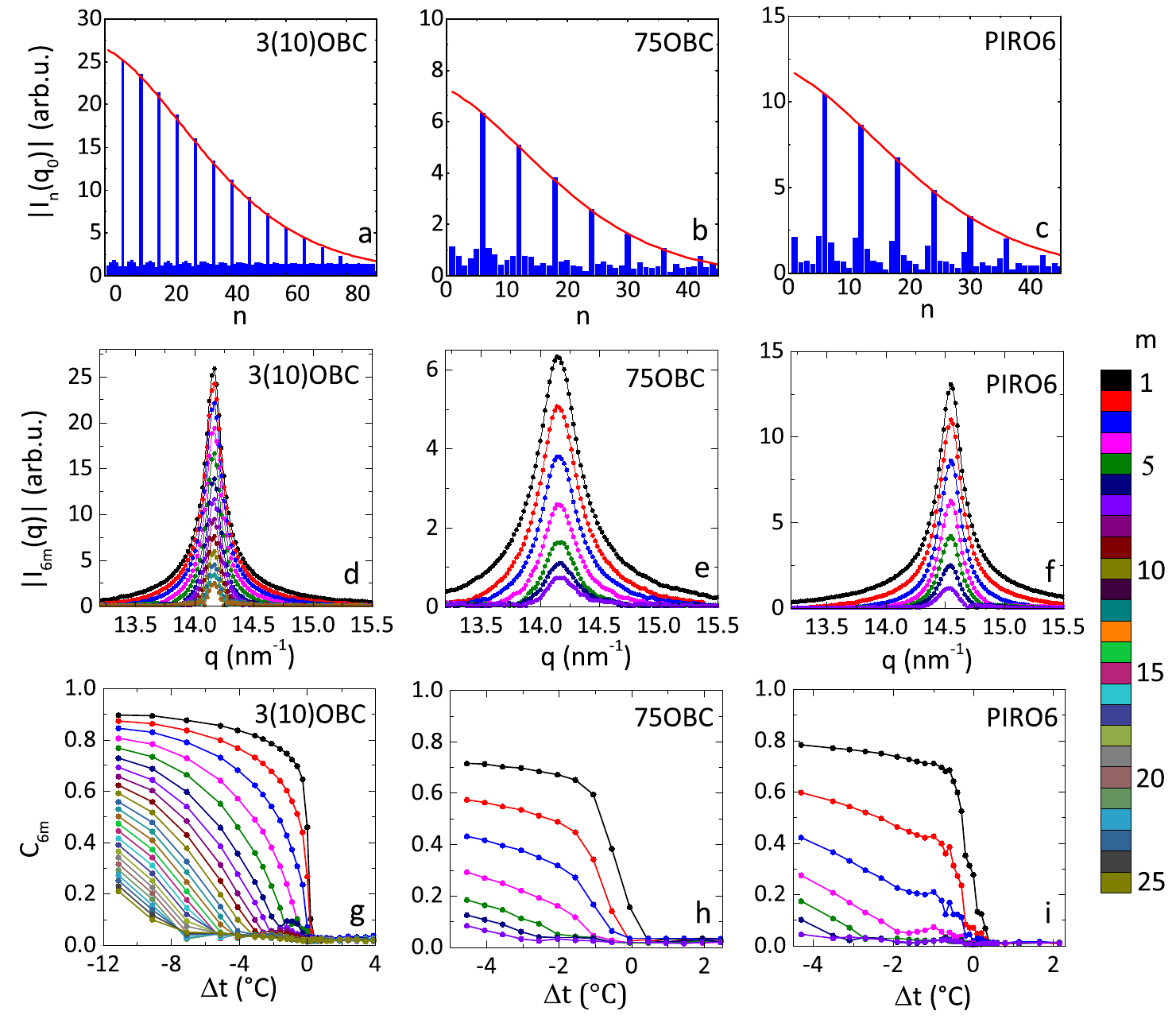}
	\caption{(a-c) Magnitudes of the Fourier components $|I_n (q_0)|$ as a function of the order number $n$ for 3(10)OBC compound at the relative temperature $\Delta t=-5.3\ ^\circ\text{C}$ (a), 75OBC compound at $\Delta t=-4.55\ ^\circ\text{C}$ (b), and PIRO6 compound at $\Delta t=-4.3\ ^\circ\text{C}$ (c). Red lines show fitting using the MCST.  (d-f) Magnitudes of the sixfold FCs of the intensity $|I_{6m}(q)|$ as a function of $q$ for 3(10)OBC (d), 75OBC (e), and PIRO6 (f) compounds at the same temperatures. (g-i) Temperature dependence of the BO order parameters $C_{6m}=|I_{6m} (q_0)/I_0 (q_0)|$ for 3(10)OBC (g), 75OBC (h), and PIRO6 (i) compounds. The experimental points are connected with solid lines for better visibility. The colorbar at the right indicates different values of $m$.}
	\label{Fig7}
\end{figure*}

For the detailed analysis of the BO order we have chosen a single-domain area of $100\times100$ $\mu$m$^2$ in our samples.
This allowed us to average the CCF over 100 diffraction patterns for 3(10)OBC and PIRO6 compounds and 25 patterns for 75OBC compound.
Due to the sixfold rotational symmetry of the diffraction patterns in the Hex-B phase only the FCs $I_n (q_0 )$ of index $n$ divisible by six should have non-zero values ($n=6,12,\ldots$), and the value of the FCs with $n\ne6,12,\ldots$ should be negligible.
Therefore, in the following we use the index $m$ defined as $n=6m$ in order to enumerate harmonics of sixfold BO order.
The magnitudes of all Fourier components $|I_n (q_0)|$ are shown in Fig.~\ref{Fig7}(a-c) as a function of the order number $n$.
It is readily seen, that the values of the FCs of indexes $n=6m$ are much higher than the values of the FCs for which $n\ne6m$.
Thus the latter FCs can be considered as a background.
The magnitudes of the FCs with $n=6m$ monotonically decrease as a function of the harmonic number $m$.
Finally, the components with $n=6m$ are counted until they reach the background level of FCs with $n\ne6m$.

The q-dependence of the FCs of intensity $|I_{6m} (q)|$ calculated on the basis of Eq.~(\ref{Eq14}) is shown for all three LC compounds in Fig.~\ref{Fig7}(d-f).
The analytical expression for the q-dependence of the FCs $|I_{6m} (q)|$ is currently unknown, this remains one of the unresolved issues of the theory.  Nevertheless, as a function of $q$ the FCs $|I_{6m} (q)|$ can be satisfactory approximated by the Lorentzian and SRL functions.
Our analysis indicated that the low order FCs ($m=1,2,3$) are better described by the SRL function. For larger values of $m$ ($m=4,5,\ldots$) the corresponding FCs are better fitted by the Lorentzian function.
The maxima of the functions $|I_{6m} (q)|$ appear at the same momentum transfer value $q=q_0$ as the maximum value of the total radial intensity $I(q)$ (see Section~\ref{sec5.1}).
An alternative approach for studying the BO order in the hexatic phase is based on the analysis of the angular shape of the diffraction peaks.
Our studies showed that at all temperatures the angular profile of the diffraction peaks in Hex-B phase can be well described by the Voigt function, which is a convolution of the Lorentzian and Gaussian functions \cite{Zaluzhnyy2017}.

\begin{figure}
	\centering
	\includegraphics[width=8.3 cm]{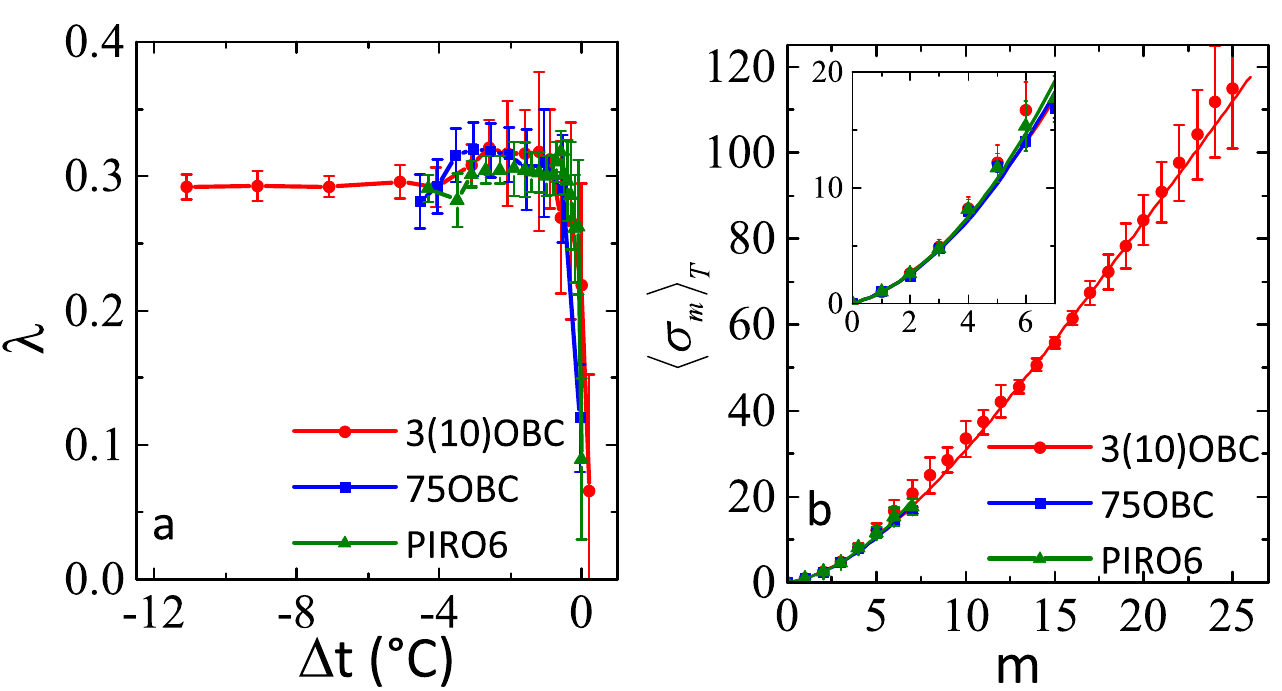}
	\caption{(a) Temperature dependence of the parameter $\lambda$, obtained from the fitting of the experimental data with the scaling relations~(\ref{Eq5}-\ref{Eq7}) for different compounds. (b) Temperature averaged values of $\langle\sigma_m\rangle_T$ and their fit with equations~(\ref{Eq6}-\ref{Eq7}) for different compounds. In the inset an enlarged region for $m=1-7$ is shown.}
	\label{Fig8}
\end{figure}

The temperature evolution of the BO order in the Hex-B phase can be described in terms of independent normalized BO order parameters introduced by Brock \textit{et al.} \cite{Brock1986,Aharony1986,Zaluzhnyy2015,Brock1989,Brock1989a}
\begin{equation}
\label{Eq16}
C_{6m}=\bigg| \frac{I_{6m}(q_0)}{I_{0}(q_0)} \bigg| \, .
\end{equation}
The number of nonzero coefficients $C_{6m}$ and their magnitudes characterize a degree of the BO order in the hexatic phase.
The temperature dependence of the BO order parameters of various order is shown for all three compounds in Fig.~\ref{Fig7}(g-i).
In the Sm-A phase all parameters $C_{6m}$ are close to zero due to absence of any angular ordering in the system.
As soon as the temperature decreases below the Sm-A -- Hex-B phase transition the BO order parameters start to appear one after another.
At the phase transition temperature only the first two or three BO order parameters can be distinguished, but during the cooling process the BO order parameters of higher order sequentially start to increase.
Although the general tendency is that the magnitudes of the BO order parameters increase during the cooling process, a non-monotonic behaviour of some of the parameters ($m=4-10$ for 3(10)OBC, $m=4-5$ for 75OBC and $m=3-5$ for PIRO6) can be observed in the temperature range, where their values are still small \cite{Zaluzhnyy2015}.

The exceptional high number ($m_{max} = 25$) of the successive BO order parameters determined  in the Hex-B phase of the 3(10)OBC can be explained by the fact that this compound exhibits the hexatic phase in a wide temperature range of about 11 $^\circ$C.
For comparison in 75OBC and PIRO6 hexatic phase exists only within the range of about 5 $^\circ$C and 6.6 $^\circ$C, respectively (see Table~\ref{Table2}).
It is clear that a higher degree of the BO order can be achieved in the Hex-B phase before its crystallization if temperature range of the hexatic phase existence is wider.
More importantly, the amount and magnitude of successive BO order parameters are influenced by the strength of interaction between the BO and positional order.
For example, in 3(10)OBC a strong correlation between the number of FCs of the BO order and the value of the positional correlation length $\xi$ has been detected \cite{Zaluzhnyy2015}.
This observation indicates strong coupling between the BO and positional order in the hexatic phase of 3(10)OBC.

A quantitative comparison of the BO parameters of different orders in the Hex-B phase was made on the basis of the MCST (see Section~\ref{sec2.3}).
At each temperature the values of $C_{6m}$ were fitted by the scaling relations~(\ref{Eq5}-\ref{Eq7}).
For the analysis of the experimental data for 3(10)OBC the approximation $x_m\approx x_m^{(1)}=\lambda-\mu m$ was used for the parameter $x_m$ in Eq.~(\ref{Eq7}).
For other samples the number of non-zero BO order parameters $C_{6m}$ was smaller, so we could not determine the value of the correction parameter $\mu$ for them.
Thus, for 75OBC and PIRO6 compounds we used approximation $x_m\approx x_m^{(0)}=\lambda$.
The temperature dependence of the parameter $\lambda$ for all three samples is shown in Fig.~\ref{Fig8}(a).
On cooling the parameter $\lambda$ rapidly grows up to the value $\lambda\approx0.3$ at the Sm-A -- Hex-B phase transition and then stays almost constant with the temperature.
The experimentally determined values of the parameters $\lambda$ and $\mu$ in the Hex-B phase are in good agreement with the predictions of the MCST for 3D hexatics (see Section~\ref{sec2.3}).

Since the exponents $\sigma_m$ in the scaling law~(\ref{Eq5}) appear to be almost independent on temperature \cite{Aharony1986,Zaluzhnyy2015}, one can calculate the temperature-averaged values $\langle\sigma_m\rangle_T$ .
In Fig.~\ref{Fig8}(b) the averaged values $\langle\sigma_m\rangle_T$ are shown as a function of the number $m$ for all three compounds.
By the fitting of the exponents $\langle\sigma_m\rangle_T$ with equations~(\ref{Eq5}-\ref{Eq7}) we obtained the following values of the MCST parameters: $\lambda=0.31$ and $\mu=0.009$ for 3(10)OBC,  $\lambda=0.27$ for 75OBC and $\lambda=0.29$ for PIRO6 (see values in Table~\ref{Table2}). We were even able to determine the magnitude of the next correction term $\nu\sim 10^{-4}$ for 3(10)OBC compound (see Eq.~(\ref{Eq7})), for which we observed the largest number of the BO order parameters \cite{Zaluzhnyy2015}.

\subsection{Critical behavior at the smectic-hexatic phase transition}
\label{sec5.4}
The critical behavior at the Sm-A -- Hex-B phase transition in LCs has been studied by various techniques for decades, but the problem is still far from being solved.
The MCST theory works for 3D hexatic films surprisingly well \cite{Brock1986,Zaluzhnyy2015,Brock1989a,Brock1989}.
This indicates that the initial assumption of the theory \cite{Aharony1986} that the Sm-A -- Hex-B transition lies in a crossover region between the mean field (tricritial behavior) and 3D XY universality class might be true.

\begin{figure*}
	\centering
	\includegraphics[width=17.1 cm]{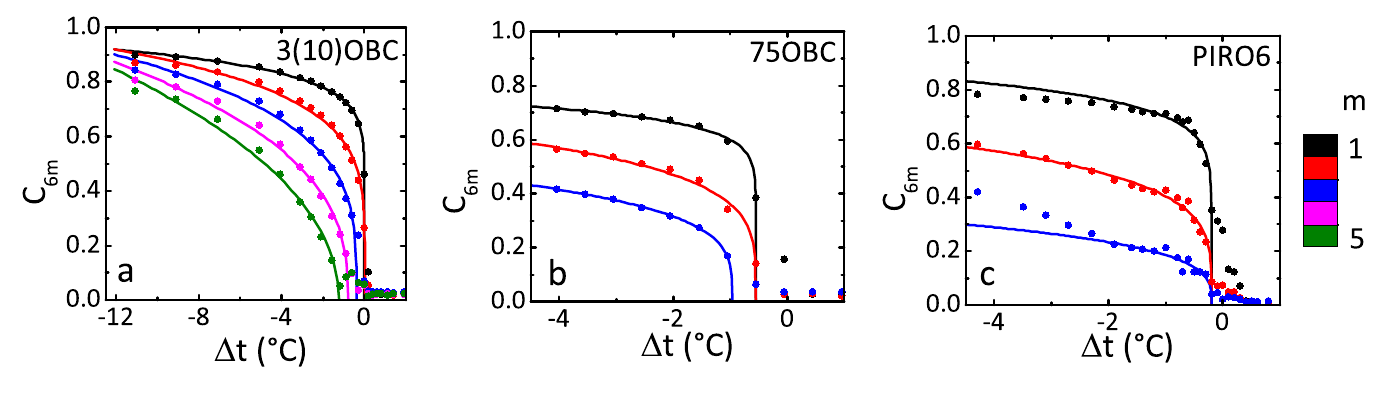}
	\caption{Temperature dependence of the BO order parameters $C_{6m}$ for compounds 3(10)OBC (a), 75OBC (b), and PIRO6 (c). Solid lines show the fitting of the experimental data by the power law~(\ref{Eq17}). The colorbar at the right indicates different values of $m$.}
	\label{Fig9}
\end{figure*}

\begin{table*}
\small
  \caption{\ Parameters of fitting of the temperature dependence of $C_{6m}$ by Eq.~(\ref{Eq17}). Typical absolute errors of parameters determination are $\Delta A_m=\pm0.01$ and $\Delta T_{C,m}=\pm0.01\,^{\circ}\text{C}$; relative error for the critical exponent is $\delta\beta_m=\pm10\%$}
  \label{Table3}
  \begin{tabular*}{\textwidth}{@{\extracolsep{\fill}}llllllllll}
    \hline
     Substance  & \multicolumn{3}{c}{3(10)OBC} & \multicolumn{3}{c}{75OBC} & \multicolumn{3}{c}{PIRO6} \\
     \cline{1-1} \cline{2-4} \cline{5-7} \cline{8-10}
     \rule{0cm}{0.3cm}
    Parameters & $A_m$ & $T_{C,m}$, $^\circ$C & $\beta_m$ & $A_m$ & $T_{C,m}$, $^\circ$C & $\beta_m$ & $A_m$ & $T_{C,m}$, $^\circ$C & $\beta_m$       \\ \hline
    m=1 & 1.06 & 66.11 & 0.09 & 0.92 & 63.25 & 0.09 & 1.12 & 92.40 & 0.10 \\

    m=2 & 1.29 & 66.11 & 0.19 & 1.07 & 63.24 & 0.22 & 1.19 & 92.40 & 0.23 \\

    m=3 & 1.45 & 65.71 & 0.27 & 0.91 & 62.84 & 0.26 & 1.45 & 92.40 & 0.29 \\

    m=4 & 1.81 & 65.44 & 0.40 &  &  &  &  &  &  \\

    m=5 & 2.00 & 64.86 & 0.48 &  &  &  &  &  &  \\
    \hline
  \end{tabular*}
\end{table*}

We propose to look at the problem of the critical behavior at the Sm-A -- Hex-B transition from a different angle. In Fig.~\ref{Fig9} the temperature dependence of the first successive BO order parameters in the vicinity of a Sm-A -- Hex-B transition is shown.
First, these curves follow the power-law with the exponents increasing with the harmonic number $m$.
Second, the higher order harmonics of the BO order appear at temperatures that are lower than the phase transition temperature $T_C$.
Taking these observations into account we fitted the temperature dependence of the BO order parameters $C_{6m}$ by a power-law
\begin{equation}
\label{Eq17}
C_{6m}=A_m\bigg|\frac{T-T_{C,m}}{T_{C,m}}\bigg|^{\beta_m} \,
\end{equation}
with three  adjustable  independent parameters: $A_m$, $T_{C,m}$ and $\beta_m$.
Here $A_m$ is a scaling constant, $T_{C,m}$ is a critical temperature, and $\beta_m$ is a critical exponent for each harmonic. The result of the fitting with Eq.~(\ref{Eq17}) is shown in Table~\ref{Table3}. The fitting gives for the critical exponent of the first harmonic $C_6$ the unusually small value, $\beta_1\approx0.1$ for all compounds.
Additionally, the higher order harmonics of the BO order scale as the powers of the first harmonic, with the exponent equal to harmonic number: $C_{12}\sim C_{6}^2$, $C_{18}\sim C_{6}^3$, etc. (see Fig.~\ref{Fig9} and~Table \ref{Table3}).

The behavior of critical exponents of higher order can be qualitatively understood in the frame of the Landau theory with nonlinear coupling between the BO order parameters of different orders.
The excess of free energy density can be expanded in the vicinity of the Sm-A -- Hex-B transition in powers of the first two hexatic order parameters $C_6$ and $C_{12}$ as
\begin{equation}
\label{Eq18}
f=\frac{1}{2}A_1C_{6}^2+\frac{1}{2}A_2C_{12}^2+\frac{1}{4}B_1C_{6}^4+\frac{1}{4}B_2C_{12}^4-D C_{6}^2C_{12} \, ,
\end{equation}
where $A_1 = a_1(T - T_{C,1})$,  $A_2 = a_2 (T - T_{C,2})$, $T_{C,2} < T_{C,1}$,  enabling the possibility for successive BO harmonics to condense at different temperatures \cite{Acknow}.
The expression~(\ref{Eq18}) contains the coupling term $-D C_{6}^2C_{12}$ of the lowest order, which forms an invariant with respect to the free energy ($C_{6}^2 \sim \exp (12i\theta)$, $C_{12}\sim \exp (12i\theta)$).
We also assume that $C_{12} < C_{6}$.

Minimization of Eq.~(\ref{Eq18}) with respect to $C_{12}$ gives the following equation
\begin{equation}
\label{Eq19}
\frac{\partial f}{\partial C_{12}}=A_2C_{12}+B_2C_{12}^3-D C_{6}^2=0 \, .
\end{equation}
Due to a small value of the parameter $C_{12}$ with respect to $C_{6}$, the second term in Eq.~(\ref{Eq19}) can be disregarded.
Finally we arrive to the simplified equation for $C_{12}$:
\begin{equation}
\label{Eq20}
C_{12}\approx\frac{D}{A_2}C_{6}^2 \, .
\end{equation}
Thus the nonlinear coupling between the BO order parameters $C_{6}$ and $C_{12}$ leads to the relation $C_{12}\sim C_{6}^2$ observed in our experiment for all three compounds (see Fig.~\ref{Fig9} and Table~\ref{Table3}).
In a similar way higher order coupling invariants of the type $C_{6}^3C_{18}$, $C^4_{6}C_{24}$, and corresponding quadratic terms can be added in the Landau expansion~(\ref{Eq18}).
We neglect here the higher order coupling terms of the type $C_{12}^3C_{18}^2$ and so on.
As a result we obtain relations $C_{18}\sim C_{6}^3$ and $C_{24}\sim C_{6}^4$ which also fit well the experimental data.

The small value of the critical exponent $\beta_1\approx0.1$ rules out the 3D XY model, which should apply to a Sm-A -- Hex-B phase transition from basic symmetry considerations.
The proximity to a tricritical point, where $\alpha=1/2$ and $\beta=1/4$, is also inconsistent with our results.
The fact that the critical exponent for specific heat $\alpha$ is not so far from 0.5 ($\alpha=0.6-0.65$ for 65OBC) might indicate that the Sm-A -- Hex-B transition occurs in the vicinity of a tricritical point as it was initially suggested by Aharony \textit{et al.} \cite{Aharony1986}.
In this case a crossover behavior can be observed, similar to the situation near the nematic -- Sm-A tricritical point \cite{Anisimov1990}.
Interestingly, our results for $\beta_1$ together with the measured specific heat critical exponent $\alpha\approx0.65$ in 65OBC conform with the 2D four-state Potts model with $\alpha=2/3$ and $\beta=1/12$ \cite{Wu1982}, which corresponds to strongly fluctuating system - three-spin model on the triangular lattice \cite{Baxter1982}.
What might be common between the 2D four-state Potts model and the hexatic BO ordering, which is described by a two-component order parameter is currently unclear.

Let us discuss the values of the critical exponents at the Sm-A -- Hex-B transition in more detail. Bruinsma and Aeppli \cite{Bruinsma1982} proposed a theory based on the coupling of BO order with a short-range  herringbone orientational order.
Due to this theory the continuous Sm-A -- Hex-B transition can be driven to first order with an appearance of the tricritical point at the transition line.
Such an approach might be relevant to the hexatics possessing low temperature Cr-E phase of the type of nmOBC series.
However, this explanation does not work for PIRO6 with a Cr-B low-temperature phase, where the short-range herringbone order is absent \cite{Gorecka1994}.

There are some other theoretical models, which introduce a tricritical point at the Sm-A -- Hex-B phase transition line.
Besides the general theory proposed by Aharony \textit{et al.} \cite{Aharony1986} (see Section~\ref{sec2.3}) there are theories by Selinger \cite{Selinger1988},  who suggested coupling between the BO order and layer displacement fluctuations, and by Haga \textit{et al.} \cite{Haga1997}, where the interaction of the BO order with the in-plane positional strain is considered.
The latter represents some variant of a compressible XY model, which allows fluctuations of the in-plane strain for weakly first-order transitions, if the material contracts on ordering
(see also compressible n-vector model \cite{Bergman1976,Moura1976}).
In principle, the renormalization of the critical exponents in the vicinity of such a tricritical point can lead to very unusual values of the effective critical indexes.
However, it is not clear how this model can be implemented to the case of the Sm-A -- Hex-B transition.

Finally, in recently published work by Kats \textit{et al.} \cite{Kats2016}, the authors have made an assumption that the translational (positional) correlation length $\xi$ is much larger than the hexatic correlation length $\xi_{h}$.
The above inequality is violated only in the narrow vicinity of the Sm-A -- Hex-B transition temperature, where the critical behavior characteristic of the superfluid phase transition is conserved.
As a result the singular part of the heat capacity represents a sum of two terms: the first with a small critical exponent $\alpha\approx-0.01$ characteristic of the 3D XY universality class and the second term with the exponent $\alpha\approx1$ which originates from the interaction of the BO order parameter with density fluctuations.
Thus, the heat capacity critical exponent $\alpha=0.6-0.65$ \cite{Huang1981,Haga1997,VanRoie2005,Mercuri2013} is based on a single, but not universal, power-law fitting of the experimental data, which reflects a crossover behavior between two regimes described above \cite{Kats2016}.

\section{Conclusions}

The important result of our study is that LC compounds of various compositions, possessing different type of low temperature crystal phase (Cr-B or Cr-E) and different range of existence of the Hex-B phase display similar thermodynamic behavior in the hexatic phase and in the vicinity of the Sm-A -- Hex-B phase transition.
For example, PIRO6, which possesses an internal (and possibly external) hydrogen bonding shows essentially the same temperature dependence of the positional correlation length, similar MCST characteristics and the same critical behavior as 3(10)OBC compound.

Utilizing a micron-sized X-ray beam we were able to study spatial variations of the BO order parameter and to construct 2D maps of the bonds orientations. We observed formation of large domains (more than 100 $\mu$m) with uniform orientation of the intermolecular bonds across the whole 3(10)OBC and PIRO6 films.
The bond angle in these domains shows continuous change over the whole area, which allows the bond angles from different domains to match themselves tangentially. This situation reminds of the formation of large scale orientational domains in nematic LCs.
For 75OBC compound the situation is different because the average bond-orientation of the hexatic domains growing from the opposite sides of the film deviate from each other. This leads to smooth variation of the bond angle $\theta(\textbf{r})$ in the transient area between the surface domains with different bond orientations.

In addition, we studied temperature dependence of the positional correlation length, which is coupled to the BO order parameters, and investigated the shape of the diffraction peaks in the radial direction. The shape is affected by large-scale BO fluctuations in the vicinity of the phase transition.
Our results support the validity of the MCST for all three compounds under study. We have confirmed experimentally the value of the first order correction term in the scaling relation of the BO order parameters that was predicted by the MCST.
We have shown that the number and magnitude of successive BO order parameters in the Hex-B phase are determined not only by the temperature range of the Hex-B phase, but primarily by the strength of interaction between the BO and positional order.

The temperature dependence of the BO order parameters in the vicinity of the Sm-A -- Hex-B transition enables direct determination of the critical exponent $\beta$ for the successive FCs.
The values of $\beta_1$ for the first harmonic of the BO order in all three compounds were found to be of the order of $\beta_1\approx0.1$.
This value of the critical exponent $\beta$ together with the large value of the heat capacity critical exponent $\alpha=0.6-0.65$ are in contradiction with the conventional 3D XY model.
The proximity to a tricritical point is also inconsistent with our results. The higher order harmonics of the BO order scale as the powers of the first harmonic, with the exponent equal to the harmonic number.
This indicates the nonlinear coupling of the BO order parameters of different orders.
Currently none of the existing theories can describe the full set of experimental data related to the Sm-A -- Hex-B transition in LCs.

Experimentally it is clear that smectic liquid crystals provide definite examples of 2D and 3D hexatic phase with the BO order.
In spite of the extensive experimental work there is no conclusive evidence for the dislocation-unbinding mechanism proposed by Halperin and Nelson \cite{Halperin1978} even for the case of 2D LC hexatic phase.
The defect-mediated theory predicts rather specific types of singularities that have not been observed in 2D hexatic LCs and in 3D case the situation becomes even more ambiguous.
For example, the theory predicts a wide bump in the specific heat above the hexatic-liquid transition point due to gradual unbinding of disclination pairs.
Instead, a nearly symmetrical specific-heat anomaly showing divergent behavior has been found at the Sm-A -- Hex-B transition in three layer \cite{Geer1993} and two layer Hex-B films \cite{Stoebe1995,Stoebe1992,Jin1996}.
As a rule the crystal-hexatic phase transition in LC hexatic films is of the first order, which preempt a continuous transition from the Hex-B to Sm-A phase \cite{Stoebe1995}.
Despite the considerable theoretical effort, there is no full understanding yet regarding the mechanism of these transitions.
Thus, the development of a microscopic theory of the hexatic phase in LCs, which necessary includes the angular ordering of the pairs of molecules forming the sides of local hexagons (molecular bonds), is in high demand.

	\begin{acknowledgments}
We acknowledge support of this project and discussions with E.~Weckert.
We are grateful to W.~H.~de~Jeu and C.~C.~Huang for supplying us with the hexatic samples and fruitful discussions.
We thank R. Gehrke for careful reading of the manuscript.
Special thanks are due to E.~I.~Kats and V.~E.~Podnek for pointing our attention to Landau theory with nonlinear coupling of order parameters and useful discussions.
We thank V.~V.~Lebedev, A.~R.~Muratov, E.~S.~Pikina and V.~M.~Kaganer for stimulating theoretical discussions. We are thankful to J.~Thoen for his comments on the specific heat measurements in 65OBC liquid crystal.
This work was partially supported by the Virtual Institute VH-VI-403 of the Helmholtz Association. The work of I.A.Z. and B.I.O. was partially supported by the Russian Science Foundation (Grant No. 14-12-00475).
	\end{acknowledgments}

\bibliography{Hex_big_ref}

\end{document}